\documentclass[a4paper]{jpconf}
\usepackage{graphicx}
\usepackage{amssymb,amsmath,amsthm}
\usepackage[all]{xy}
\xyoption{curve}
\xyoption{line}
\usepackage{graphicx}

\def\one{\mbox{1 \kern-.59em {\rm l}}}

\def\alg{{\mathcal A}}
\def\balg{{\mathcal B}}
\def\calg{{\mathcal C}}
\def\dalg{{\mathcal D}}

\def\hil{{\mathcal H}}
\def\bun{{\mathcal E}}
\def\Poin{{\mathcal P}}

\def\Dirac{{D\!\!\!\!/\,}} 

\def\Ind{{\rm index}}
\def\Hom{{\rm Hom}}
\def\Ext{{\rm Ext}}
\def\End{{\rm End}}
\def\K{{\rm K}}
\def\KK{{\rm KK}}

\def\H{{\rm H}}
\def\P{{\rm P}}

\def\HP{{\H\P}}

\def\HL{{{\rm HL}}}
\def\op{{\rm o}}
\def\Id{{\rm id}}
\def\pt{{\rm pt}}
\def\Cl{{\rm Cliff}}
\def\ch{{\rm ch}}

\def\Pic{{\rm Pic}}

\def\Todd{{\rm Todd}}

\def\bb#1{\hbox{\mybb#1}}

\def\nn{\nonumber}

\def\be{\begin{equation}}
\def\ee{\end{equation}}
\def\bea{\begin{eqnarray}}
\def\eea{\end{eqnarray}}
\def\bd{\begin{displaymath}}
\def\ed{\end{displaymath}}

\def\dd{{\rm d}}
\def\bb{{\rm b}}

\newcommand{\R}{{\mathbb{R}}}

\newcommand{\C}{{\mathbb{C}}}
\newcommand{\Z}{{\mathbb{Z}}}
\newcommand{\Q}{{\mathbb{Q}}}
\newcommand{\torus}{{\mathbb{T}}}
\newcommand{\dtorus}{{\widehat{\torus}{}}}

\def\ms{\mathfrak{s}}

\begin{document}

\begin{flushright}
HWM--07--35\\
EMPG--07--18\\
\end{flushright}

\title{D-branes, KK-theory and \\ duality on noncommutative
  spaces\footnote{Based on invited talks given by R.J.S. at the
    International Conferences ``Noncommutative Spacetime Geometries'',
    March~26--31, 2007, Alessandria, Italy, and ``Noncommutative
    Geometry and Physics'', April~23--27, 2007, Orsay, France. To be
    published in {\sl Journal of Physics Conference Series}.}}

\author{J.~Brodzki$^{(a)}$, V.~Mathai$^{(b)}$, J.~Rosenberg$^{(c)}$ and
  R.J.~Szabo$^{(d)}$}

\address{$^{(a)}$ School of Mathematics, University of Southampton\\
Southampton SO17 1BJ, UK}
\address{$^{(b)}$ Department of Pure Mathematics, University of
  Adelaide\\ Adelaide 5005, Australia}
\address{$^{(c)}$ Department of Mathematics, University of Maryland\\
College Park, MD 20742, USA}
\address{$^{(d)}$ Department of Mathematics and \\ Maxwell
  Institute for Mathematical Sciences \\ Heriot-Watt University,
Riccarton, Edinburgh EH14 4AS, UK}

\ead{j.brodzki@soton.ac.uk , mathai.varghese@adelaide.edu.au ,
  jmr@math.umd.edu R.J.Szabo@ma.hw.ac.uk}

\begin{abstract}
We present a new categorical classification framework for D-brane
charges on noncommutative manifolds using methods of bivariant
K-theory. We describe several applications including an explicit
formula for D-brane charge in cyclic homology, a refinement of open
string T-duality, and a general criterion for cancellation of global
worldsheet anomalies.
\end{abstract}

\setcounter{equation}{0}
\section{Introduction\label{Intro}}

This article centres in part around the following physical question:
\emph{What is a D-brane?} More precisely, given a closed string
background $X$, what are the possible states of D-branes in $X$? At
the level of the worldsheet field theory of open strings in $X$, this
is a problem of finding the consistent boundary conditions in the
underlying boundary conformal field theory. As we will discuss, when
this worldsheet perspective is combined with the target space
classification of D-branes in terms of Fredholm modules over a suitable
$C^*$-algebra, a powerful categorical description of D-branes and
their charges emerges. This is particularly useful for those boundary
states which have no geometric
description. In certain instances these ``non-geometric'' backgrounds
can be interpreted as noncommutative manifolds, \emph{i.e.}, as
separable noncommutative $C^*$-algebras. The formalism that we
review in the following was developed in detail in
refs.~\cite{BMRS1,BMRS2}, and it allows for the construction of general
charge vectors for D-branes on these noncommutative spaces.

This point of view becomes particularly fruitful for considerations
involving compactifications with $H$-flux. Consider, for example, a
principal torus bundle $X\to M$ with constant $H$-flux. Applying a
T-duality transformation along the fibre gives a space which does not
always admit a global riemannian description. Instead, one can double
the dualized directions and use elements of the T-duality group as
transition functions between local patches. This is called a
``T-fold''~\cite{Hull:2004in}. In some examples, one can
show~\cite{Grange:2006es} that the \emph{open} string metric on a
T-fold is precisely the metric on an associated continuous field of
stabilized noncommutative tori fibred over $M$ which corresponds to a
certain crossed product $C^*$-algebra~\cite{BEM,MR}. Thus the open
string version of a T-fold can be generally regarded as a globally
defined, noncommutative $C^*$-algebra. Performing additional T-duality
transformations along the base leads in some instances to
nonassociative tori in the fibre
directions~\cite{BHM2}--\cite{EH}. This example, wherein the action of
T-duality is realized by taking a certain crossed product algebra,
motivates an axiomatic definition of topological open string
T-duality. This generalizes and refines the more common examples of
T-duality between noncommutative spaces in terms of Morita
equivalence~\cite{SW} to a special type of ``KK-equivalence'', which
defines a T-duality action that is of order two up to Morita
equivalence.

From a purely mathematical perspective of noncommutative geometry, the
framework needed to achieve the physical constructions above develops
more tools for dealing with noncommutative spaces in general. These
include the appropriate noncommutative versions of Poincar\'e duality
and orientation, topological invariants of noncommutative spaces such
as the Todd genus, and a noncommutative version of the
Grothendieck-Riemann-Roch theorem which is directly linked to the
general formula for D-brane charge. All of this is defined and
developed in the purely algebraic framework of separable
$C^*$-algebras.

\setcounter{equation}{0}
\section{D-branes and K-theory\label{DbranesK}}

It is well-known that D-brane charges and Ramond-Ramond fields in
Type~II superstring theory without $H$-flux are classified
topologically by the complex K-theory of spacetime
$X$~\cite{MM}--\cite{WOverview}. We will begin by briefly reviewing
some salient features of this classification that we will generalize
later on to more generic noncommutative settings.

\subsection{A simple observation}

Let $X$ be a compact spin$^c$ manifold. Poincar\'e duality in
cohomology states that the natural bilinear pairing
\be
(x,y)_\H=\big\langle x\,\smile\,y\,,\,[X]\big\rangle
\label{cohpairing}\ee
between cohomology classes $x,y$ of $X$ in complementary degree is
non-degenerate. If $\alpha,\beta$ are de~Rham representatives of
$x,y$, then this pairing is just
$(x,y)_\H=\int_X\,\alpha\wedge\beta$. On the other hand, in K-theory
the natural bilinear pairing between complex vector bundles $E,F\to X$
is given by the index of the twisted Dirac operator
\be
(E,F)_\K=\Ind\big(\Dirac_{E\otimes F}\big)
\label{Kpairing}\ee
associated to the spin$^c$ structure on $X$. The Chern character gives
a natural, $\Z_2$-graded ring isomorphism
\be
\ch\,:\,\K(X)\otimes\Q~\xrightarrow{\approx}~\H(X,\Q)
\label{Cherniso}\ee
but it doesn't preserve these bilinear forms. However, by the
Atiyah-Singer index theorem one has
\be
\Ind\big(\Dirac_{E\otimes F}\big)=\big\langle\ch(E\otimes F)\,
\smile\,\Todd(X)\,,\,[X]\big\rangle \ ,
\label{ASindexthm}\ee
so we get an isometry by replacing the isomorphism (\ref{Cherniso})
with the ``twisted'' Chern character
\be
\ch~\longrightarrow~\sqrt{\Todd(X)}\,\smile\,\ch \ .
\label{twistedChern}\ee
Here $\Todd(X)\in\H(X,\Q)$ is the invertible Todd characteristic class
of the tangent bundle of $X$, which can be expressed in terms of the
Pontrjagin classes of $X$ along with a degree two characteristic class
$c_1\in\H^2(X,\Z)$ whose reduction modulo~$2$ is the second
Stiefel-Whitney class $w_2(X)$. This almost trivial observation plays
a crucial role in what follows.

\subsection{D-brane charges}

A natural geometric description of a D-brane in $X$ is provided by a
topological K-cycle $(W,E,f)$ in $X$~\cite{Harvey:2000te}--\cite{RSV},
where $f:W\hookrightarrow X$ is a closed, embedded spin$^c$
submanifold of $X$ (the brane worldvolume), and $E\to W$ is the
Chan-Paton gauge bundle equipped with a hermitean connection and
regarded as an element of the topological K-theory group
$\K^0(W)$. The collection of K-cycles forms an additive category under
disjoint union. The quotient of this category of D-branes by
Baum-Douglas ``gauge'' equivalence~\cite{BD} is isomorphic to the
K-homology of $X$, defined as the group of stable homotopy classes of
\emph{Fredholm modules} over the commutative $C^*$-algebra $\alg=C(X)$
of continuous functions on $X$. The isomorphism is generated by
associating to a K-cycle $(W,E,f)$ the (unbounded) Fredholm module
$(\hil,\rho,\Dirac_E^{(W)})$, where $\hil=L^2(W,S\otimes E)$ with
$S\to W$ the spinor bundle over the D-brane worldvolume, the
$*$-representation $\rho(\phi)=m_{\phi\circ f}$ of $\phi\in\alg$ on
the separable Hilbert space $\hil$ is given by pointwise
multiplication by the function $\phi\circ f$, and $\Dirac_E^{(W)}$ is
the Dirac operator associated to the spin$^c$ structure on $W$.

It follows that D-branes naturally provide K-homology classes on
$X$. They are dual to K-theory classes $f_!(E)\in\K^d(X)$, where $f_!$
is the K-theoretic Gysin pushforward map and $d$ is the codimension of
$W$ in $X$. The \emph{Ramond-Ramond charge} of a D-brane supported on
$W$ with Chan-Paton bundle $E\in\K^0(W)$ is the element of $\H(X,\Q)$
given by
\be
Q(W,E)=\ch\big(f_!(E)\big)\,\smile\,\sqrt{\Todd(X)} \ .
\label{MMformula}\ee
This is known as the \emph{Minasian-Moore formula}~\cite{MM}. One of
our goals in the following will be to generalize this construction to
generic noncommutative settings.

\setcounter{equation}{0}
\section{D-branes and bivariant K-theory}

We will now propose a powerful categorical classification of D-branes
which extends the descriptions provided by K-theory and K-homology in
a unified manner. Our proposal is motivated by the structure of the
open string algebras and bimodules that arise in the underlying
worldsheet boundary conformal field theory, which enable us to treat
the collection of allowed D-brane boundary conditions as a certain
category.

\subsection{D-brane categories}

Open string fields define relative maps
$(\Sigma,\partial\Sigma)\to(X,W)$ from an oriented Riemann surface
$\Sigma$ with boundary $\partial\Sigma$. Not all maps are
allowed. They are constrained by the requirements of worldsheet
conformal and modular invariance (such as the Cardy conditions), as
well as by cancellation of global worldsheet anomalies. These
constraints are viewed as equations of motion in the underlying
boundary conformal field theory. For example, in Type~II superstring
theory in the absence of $H$-flux, this is just our previous
requirement that the worldvolume $W$ be a spin$^c$
manifold. Classically, this means that a D-brane may be regarded as a
suitable boundary condition in the boundary conformal field theory. It
is not presently known what is meant generally by a ``quantum
D-brane'' in the underlying quantum boundary conformal field
theory. In the following we will propose an algebraic characterization
of quantum D-branes in the context of separable $C^*$-algebras.

The crucial observation is that the concatenation of open string
vertex operators defines algebras and bimodules. We take
$\Sigma=\R\times I$, where $\R$ parametrizes the time evolution
and the interval $I=[0,1]$ parametrizes the space coordinates of the
open strings in $X$. Let us label the allowed D-brane boundary
conditions by $a,b,\dots$. An $a$-$b$ open string has $a$ boundary
conditions at its $t=0$ end and $b$ boundary conditions at its $t=1$
end. The set of $a$-$a$ open strings forms a noncommutative algebra
$\dalg_a$ of open string fields as the vertex operator algebra of
observables in the boundary conformal field theory. The opposite
algebra $\dalg_a^\op$, \emph{i.e.}, the algebra with the same
underlying vector space as $\dalg_a$ but with the product reversed, is
obtained by reversing the orientations of the $a$-$a$ open
strings. The set of $a$-$b$ boundary conditions, on the other hand,
forms a $\dalg_a$-$\dalg_b$ bimodule $\bun_{ab}$. The dual bimodule
$\bun_{ab}^\vee=\bun_{ba}$ is obtained by reversing the orientations
of the $a$-$b$ open strings. Note that $\bun_{aa}=\dalg_a$ is the
trivial $\dalg_a$-bimodule obtained by letting $\dalg_a$ act on itself
by multiplication from the left and from the right.

We would now like to define an additive category whose objects are the
D-brane algebras $\dalg_a$, and the morphisms between any two objects
$\dalg_a,\dalg_b$ is precisely the open string bimodule
$\bun_{ab}$. This means that for any three boundary conditions $a,b,c$
there should be a $\C$-bilinear map
\be
\bun_{ab}\times\bun_{bc}~\longrightarrow~\bun_{ac}
\label{bunassmap}\ee
which defines the associative composition law in this category. A
natural guess for this map is the canonical open string vertex which
combines an $a$-$b$ open string field with a $b$-$c$ open string field
into an $a$-$c$ open string field. However, the operator product
expansion on the underlying open string vertex operator algebras is not
generally associative, and in general does not even lead to a
well-defined map (\ref{bunassmap})~\cite{BMRS2}. We therefore need
some other way to define and compose the morphisms between the objects
$\dalg_a$ of our category.

\subsection{Seiberg-Witten limit\label{SWlimit}}

An example of a situation in which the assignment (\ref{bunassmap}) is
well-defined was worked out by Seiberg and Witten~\cite{SW} (see
also~ref.~\cite{WOverview}) in the case of open string boundary
conditions of maximal support on an $n$-torus $X=\mathbb{T}^n$ with a
constant $B$-field. The Seiberg-Witten limit of the boundary conformal
field theory in $X$ amounts to simultaneously sending both the string
tension $T$ and the $B$-field to infinity whilst keeping their ratio
$B/T$ a finite constant. One also needs to scale the closed string
metric $g$ to~$0$. This low-energy limit keeps only zero modes of the
string fields. Quantization of the point particle at the endpoint of
an $a$-$a$ open string gives a Hilbert space $\hil_a$ which is a
module for the noncommutative $C^*$-algebra of a noncommutative torus
$\dalg_a$. The algebra $\dalg_a\otimes\dalg_b$ acts irreducibly on the
Hilbert space $\bun_{ab}=\hil_a\otimes\hil_b^\vee$, and the map
(\ref{bunassmap}) in this case is given by
\be
V_{ac}(t'\,)=\lim_{t\to t'}\,V_{ab}(t)\cdot V_{bc}(t'\,) \ .
\label{SWOPE}\ee
Here $V_{ab}(t)$, $t\in I$ are the open string vertex operators for
the boundary conditions labelled by $a,b$, and the product in
eq.~(\ref{SWOPE}) is the operator product expansion taken in the
Seiberg-Witten limit. The map (\ref{bunassmap}) is now
well-defined as the conformal dimensions of all vertex operators,
being proportional to $g/T$, vanish in the limit~\cite{BMRS2}. In
addition, the operator product expansion (\ref{SWOPE}) is associative
in the limit, and hence the map (\ref{bunassmap}) extends to a map
\be
\bun_{ab}\otimes_{\dalg_b}\bun_{bc}~\longrightarrow~\bun_{ac} \ .
\label{bunassmapSW}\ee
Because the noncommutative algebras $\dalg_a$ contain the complete set
of observables for boundary conditions of maximal support, they act
irreducibly on the quantum mechanical Hilbert spaces and there are
natural identifications
\be
\dalg_a~\cong~\bun_{ab}\otimes_{\dalg_b}\bun_{ba} \qquad \mbox{and}
\qquad \dalg_b~\cong~\bun_{ba}\otimes_{\dalg_a}\bun_{ab} \ .
\label{dalgids}\ee
These relations mean that the open string bimodule $\bun_{ab}$ is a
\emph{Morita equivalence bimodule}, expressing a \emph{T-duality}
between the noncommutative tori $\dalg_a$ and
$\dalg_b$~\cite{SW,WOverview}.

\subsection{KK-theory}

Motivated by the situation described by the Seiberg-Witten limit of
boundary conformal field theory, we will assume that there is a
suitable extension or ``deformation'' of the open string bimodule
$\bun_{ab}$ to a \emph{Kasparov bimodule} $(\bun_{ab},F_{ab})$. In the
Seiberg-Witten limit described in Section~\ref{SWlimit} above,
\emph{i.e.}, when $\bun_{ab}$ is a Morita equivalence bimodule, this
is a ``trivial'' bimodule $(\bun_{ab},0)$. The Kasparov bimodules
$(\bun_{ab},F_{ab})$ generalize Fredholm modules and their stable
homotopy classes define the $\Z_2$-graded \emph{bivariant K-theory} or
\emph{KK-theory} group $\KK_\bullet(\dalg_a,\dalg_b)$. Elements of
this group may be regarded as ``generalized'' morphisms
$\dalg_a\to\dalg_b$ between separable $C^*$-algebras. More precisely,
there is an additive category whose objects are separable
$C^*$-algebras $\alg$ and whose morphisms between two objects
$\alg,\balg$ are exactly the elements of $\KK_\bullet(\alg,\balg)$. In
particular, if $\phi:\alg\to\balg$ is $*$-homomorphism of
$C^*$-algebras, then there is a canonically defined element
$[\phi]\in\KK(\alg,\balg)$ which is represented by the ``Morita-type''
bimodule $(\balg,\phi,0)$. This categorical point of view enables one
to uniquely characterize the bivariant K-theory groups by the
properties of homotopy invariance, stability under compact
perturbation, and split exactness on the category of separable
$C^*$-algebras and $*$-homomorphisms~\cite{Higson}.

The KK-category is the one we shall take as our toy model for a
``category of D-branes''. It is not an abelian category, but it admits
the structure of a triangulated category~\cite{Nest} and hence has
properties similar to the more commonly used categories of topological
D-branes~\cite{Aspinwall:2004jr,Moore:2006dw}. We assume that, at
least under certain circumstances, it is the appropriate category for
dealing with quantum D-branes for energies outside the classical
regime of the infinite tension limit, along the lines suggested in
ref.~\cite{WOverview}. Further properties of this category are
discussed below. The groups $\KK_\bullet(\alg,\balg)$ unify the
K-theory and K-homology of $C^*$-algebras, which arise as special
cases. When $\alg=\C$ the group
$\KK_\bullet(\C,\balg)=\K_\bullet(\balg)$ is the K-theory of
$\balg$. On the other hand, when $\balg=\C$ the group
$\KK_\bullet(\alg,\C)=\K^\bullet(\alg)$ is the K-homology of $\alg$,
as in this case a Kasparov bimodule is the same thing as a Fredholm
module over the algebra $\alg$.

\subsection{Intersection product}

Although the groups $\KK_\bullet(\alg,\balg)$ naturally incorporate
both the K-theory and K-homology classifications of D-branes, the
bivariant version of K-theory is much more powerful than K-theory or
K-homology alone. This is due to the existence of the bilinear,
associative
\emph{intersection product}
\be
\otimes_{\balg}\,:\,\KK_i(\alg,\balg)\times\KK_j(\balg,\calg)~
\longrightarrow~\KK_{i+j}(\alg,\calg) \ .
\label{intprod}\ee
The definition of this product is notoriously difficult. In
Section~\ref{T-duality} we will see how to describe it explicitly on
the category of smooth manifolds. The product (\ref{intprod}) is
compatible with composition of morphisms. If $\phi:\alg\to\balg$ and
$\psi:\balg\to\calg$ are $*$-homomorphisms of separable
$C^*$-algebras, then
\be
[\phi]\otimes_\balg[\psi]=[\psi\circ\phi] \ .
\label{compmor}\ee
The intersection product makes the group $\KK_0(\alg,\alg)$ into a
ring with unit $1_\alg=[\Id_\alg]$.

Any fixed element $\alpha\in\KK_d(\alg,\balg)$ determines
homomorphisms in K-theory and K-homology by left and right
multiplication
\be
\otimes_\alg\alpha\,:\,\K_j(\alg)~\longrightarrow~\K_{j+d}(\balg)
\qquad \mbox{and} \qquad
\alpha\otimes_\balg\,:\,\K^j(\balg)~\longrightarrow~\K^{j+d}(\alg) \ .
\label{alphahom}\ee
If $\alpha$ is \emph{invertible}, \emph{i.e.}, if there exists an
element $\beta\in\KK_{-d}(\balg,\alg)$ such that
  $\alpha\otimes_\balg\beta=1_\alg$ and
  $\beta\otimes_\alg\alpha=1_\balg$, then the maps (\ref{alphahom})
  induce isomorphisms
\be
\K_j(\alg)~\cong~\K_{j+d}(\balg) \qquad \mbox{and} \qquad
\K^j(\balg)~\cong~\K^{j+d}(\alg) \ .
\label{alphaiso}\ee
In this case the algebras $\alg$ and $\balg$ are said to be
\emph{KK-equivalent}. For example, by eq.~(\ref{dalgids}) it follows
that a Morita equivalence implies a KK-equivalence with invertible
class $\alpha=[(\bun_{ab},0)]$. The converse, however, is not
generally true.

The Kasparov intersection product defines the associative composition
law in the KK-category. It also yields the additional structure of a
tensor category with multiplication bifunctor given by the spatial
tensor product on objects, the external Kasparov product on morphisms,
and with identity object the one-dimensional $C^*$-algebra $\C$. More
precisely, this data defines a ``relaxed'' or ``weak'' monoidal
category whereby associativity of the tensor product only holds up to
natural isomorphism. A diagrammatic calculus in this tensor category
was introduced in ref.~\cite{BMRS1} and developed more extensively in
ref.~\cite{BMRS2}.

\setcounter{equation}{0}
\section{Duality and worldsheet anomaly cancellation}

We will now apply KK-theory to formulate the notion of noncommutative
Poincar\'e duality, introduced originally by
Connes~\cite{ConnesBook}. This can be applied to formulate target
space consistency conditions on noncommutative D-branes represented by
generic separable $C^*$-algebras, and hence it selects the consistent
sets of D-branes from the KK-category. Moreover, it implies that the
K-theory and K-homology classifications of D-branes are equivalent.

\subsection{Poincar\'e duality}

Let $\alg$ be a separable $C^*$-algebra and $\alg^\op$ its opposite
algebra. We say that $\alg$ is a \emph{Poincar\'e duality} (\emph{PD})
\emph{algebra} if there exists a \emph{fundamental class}
$\Delta\in\KK_d(\alg\otimes\alg^\op,\C)=\K^d(\alg\otimes\alg^\op)$ in
K-homology with an inverse class
$\Delta^\vee\in\KK_{-d}(\C,\alg\otimes\alg^\op)=
\K_{-d}(\alg\otimes\alg^\op)$ in K-theory such that
\bea
\Delta^\vee\otimes_{\alg^\op}\Delta&=&1_\alg\in\KK_0(\alg,\alg) \ ,
\nn\\[4pt]
\Delta^\vee\otimes_{\alg}\Delta&=&(-1)^d~
1_{\alg^\op}\in\KK_0(\alg^\op,\alg^\op) \ .
\label{fundclassdef}\eea
The opposite algebra is used in this definition to describe
$\alg$-bimodules as $(\alg\otimes\alg^\op)$-modules, and the sign in
eq.~(\ref{fundclassdef}) depends on the orientation of the Bott
element. This data determines inverse isomorphisms
\be
\K_i(\alg)~\xrightarrow{\otimes_\alg\Delta}~\K^{i+d}(\alg^\op)=
\K^{i+d}(\alg) \qquad \mbox{and} \qquad
\K^i(\alg)=\K^{i}(\alg^\op)~
\xrightarrow{\Delta^\vee\otimes_{\alg^\op}}~
\K_{i-d}(\alg)
\label{inverseisos}\ee
between the K-theory and K-homology of the algebra $\alg$. More
generally, by replacing $\alg^\op$ everywhere in the above by another
separable $C^*$-algebra $\balg$ gives the notion of \emph{PD pairs}
$(\alg,\balg)$. The moduli space of fundamental classes of a given
algebra $\alg$ is isomorphic to the group of invertible elements in
the unital ring $\KK_0(\alg,\alg)$~\cite{BMRS1}. This space is in
general larger than the set of all K-orientations or ``spin$^c$
structures'' discussed below.

A simple example of a PD pair $(\alg,\balg)$ is provided by taking
$\alg=C_0(X)$ to be the algebra of continuous functions vanishing at
infinity on a complete oriented manifold $X$, and either
$\balg=C_0(T^*X)$ or $\balg=C_0(X,\Cl(T^*X))$ where $\Cl(T^*X)$ is the
Clifford algebra bundle of the cotangent bundle over $X$. The
fundamental class $\Delta$ in this case is given by the Dirac operator
constructed on $\Cl(T^*X)$. When $X$ is spin$^c$, $\alg$ is itself a
PD algebra with fundamental class $\Delta$ the spin$^c$ Dirac operator
$\Dirac$ induced on the diagonal of $X\times X$, and $\Delta^\vee$ is
the Bott element. The two-dimensional noncommutative tori
$\torus_\theta^2$ are examples of noncommutative PD
algebras~\cite{BMRS1,ConnesBook}.

\subsection{K-orientation}

Suppose that $f:\alg\to\balg$ is a $*$-homomorphism of separable
$C^*$-algebras in a suitable category. Then a \emph{K-orientation} for
$f$ is a functorial way of associating an element
$f!\in\KK_d(\balg,\alg)$. This determines a \emph{Gysin} or
``\emph{wrong way}'' \emph{homomorphism} by right multiplication
\be
f_!=\otimes_\balg f!\,:\,\K_\bullet(\balg)~\longrightarrow~
\K_{\bullet+d}(\alg) \ .
\label{wrongwaymap}\ee
If $\alg$ and $\balg$ are both PD algebras, then any morphism
$f:\alg\to\balg$ is K-oriented with K-orientation
\be
f!=(-1)^{d_\alg}~\Delta_\alg^\vee\otimes_{\alg^\op}[f^\op]
\otimes_{\balg^\op}\Delta_\balg
\label{GysinPD}\ee
and $d=d_\alg-d_\balg$. We have used the fact that the involution
$\alg\to\alg^\op$ on the stable homotopy category of $C^*$-algebras
passes to the KK-category, and $[f^\op]$ is the KK-class of the
$*$-homomorphism $f^\op:\alg^\op\to\balg^\op$ defined by
$f^\op(x^\op)=f(x)^\op$ for $x\in\alg$. The functoriality of the
construction (\ref{GysinPD}), \emph{i.e.}, that $g!\otimes_\balg
f!=(g\circ f)!$ for any other morphism $g:\balg\to\calg$ of PD
algebras, follows by associativity of the Kasparov intersection
product.

For example, let $f:W\hookrightarrow X$ be a smooth proper embedding
of codimension $d$ between smooth compact manifolds such that the
normal bundle $\nu=f^*(TX)/TW$ over $W$ in $X$ is spin$^c$. Then a
K-orientation for $f$ is determined by the element
\be
f!=i^W!\otimes_{C_0(\nu)}j!
\label{Koremb}\ee
of $\KK_d(C(W),C(X))$, where
$i^W!$ is the invertible element of the KK-theory group
$\KK_d(C(W),C_0(\nu))$ determined by the Thom isomorphism of the zero
section embedding $i^W:W\hookrightarrow\nu$, and $j!$ is the element
of $\KK_0(C_0(\nu),C(X))$ induced by the extension by zero. When $X$
is spin$^c$, the spin$^c$ condition on $\nu$ is equivalent to a
spin$^c$ structure on $W$ and is just the Freed-Witten anomaly
cancellation condition for a D-brane supported on $W$ in Type~II
superstring theory on $X$ without $H$-flux~\cite{FW}. Thus any D-brane
$(W,E,f)$ in $X$ determines a canonically defined KK-theory element
$f!\in\KK(C(W),C(X))$. Our notion of K-orientation may be regarded in
this way as a generalization of the Freed-Witten condition to more
general (noncommutative) spacetime geometries and D-branes. For
example, the construction of the K-orientation (\ref{Koremb}) can be
extended to arbitrary smooth proper maps $f:W\to X$ between smooth
manifolds for which the bundle $TW\oplus f^*(TX)$ over $W$ is
spin$^c$~\cite{BMRS2}. Again when $X$ itself is spin$^c$, so is $W$
and this corresponds to a D-brane wrapping a generally
non-representable cycle in $X$ associated to a generic Baum-Douglas
K-cycle~\cite{ReisSz}.

\setcounter{equation}{0}
\section{Open string T-duality\label{T-duality}}

We will now apply our considerations thus far to give a very general,
axiomatic description of T-duality in open string theory which refines
the usual notions of topological T-duality. The formulation may be
motivated by the correspondence picture for KK-theory, introduced
originally for the KK-theory of manifolds by Baum, Connes and
Skandalis~\cite{CS}, which provides an explicit description of the
intersection product and of the KK-category itself.

\subsection{Correspondences}

Let $X,Y$ be smooth manifolds. A \emph{correspondence} is given by a
diagram
\be
\xymatrix{ &(Z,E)\ar[ld]_f\ar[rd]^g& \\ X & & Y }
\label{corrgen}\ee
where $Z$ is a smooth manifold, $E$ is a complex vector bundle over
$Z$, $f:Z\to X$ is smooth and proper, and $g:Z\to Y$ is
K-oriented. Any correspondence naturally defines a class
$g_!(f^*(-)\otimes E)$ in the bivariant K-theory group
$\KK(X,Y):=\KK(C_0(X),C_0(Y))$. This gives a geometrical realization
of the analytic index for families of elliptic operators on $X$
parametrized by $Y$. The collection of all correspondences for $X,Y$
forms an additive category under disjoint union. The quotient of this
category by the suitable notions of cobordism, direct sum and vector
bundle modification is isomorphic to the KK-theory group
$\KK(X,Y)$~\cite{BMRS2}. When $Y=\pt$, this definition reduces to the
Baum-Douglas K-homology of $X$. When $X=\pt$, it mimicks the
Atiyah-Bott-Shapiro construction of D-brane charge as an element of
the K-theory of spacetime $Y$~\cite{W,OS99,ReisSz}.

A major advantage of the correspondence picture is that the
intersection product on KK-theory, which is difficult to define in the
analytic setting, is particularly simple in this geometric setting. It
is the bilinear associative map
\be
\otimes_M\,:\,\KK(X,M)\times\KK(M,Y)~\longrightarrow~
\KK(X,Y)
\label{intprodM}\ee
which is defined by sending two correspondences
\be
\xymatrix{ &(Z_1,E_1)\ar[ld]_f\ar[rd]^{g_M}& &
(Z_2,E_2)\ar[ld]_{f_M}\ar[rd]^{g} & \\ X & & M & & Y }
\label{2corrs}\ee
to the correspondence
\be
[Z,E]=[Z_1,E_1]\otimes_M[Z_2,E_2]
\label{corrprod}\ee
with $Z=Z_1\times_M Z_2$ and $E=E_1\boxtimes E_2$. This definition
requires a transversality condition on the two maps $f_M$ and $g_M$ in
order to ensure that the fibred product $Z$ is a smooth manifold.

\subsection{Fourier-Mukai transform\label{FourierMukai}}

While the correspondence picture is mathematically useful because it
gives a somewhat more precise meaning to the interpretation of
KK-theory classes as ``generalized morphisms'', its main physical
appeal is that it strongly resembles a smooth analog of the
Fourier-Mukai transform and is thus intimately related to
T-duality~\cite{Hori}. We will now make this observation precise. Let
$M$ be a smooth manifold, and let $\torus^n\cong\R^n/\Z^n$ be an
$n$-dimensional torus. Let $\widehat{\torus}{}^n$ be the dual
$n$-torus, which is canonically isomorphic to the Picard group
$\Pic^0(\torus^n)$ of flat line bundles over $\torus^n$. The
Poincar\'e line bundle is the unique line bundle $\Poin_0$ over the
product $\torus^n\times\dtorus{}^n$ such that for any point
$\widehat{t}\in\dtorus{}^n$ the restriction $(\Poin_0)_{\widehat{t}}$ to
$\torus^n\times\{\,\widehat{t}~\}$ represents the element of
$\Pic^0(\torus^n)$ corresponding to $\widehat{t}$, and such that the
restriction bundle $\Poin_0|_{\{0\}\times\dtorus{}^n}$ is trivial.

Consider the diagram
\be
\xymatrix{ &\big(M\times\torus^n\times\dtorus{}^n\,,\,\Poin\big)
\ar[ld]_{p_1}\ar[rd]^{p_2}& \\ M\times\torus^n & & M\times\dtorus{}^n }
\label{FMcorr}\ee
where $p_1,p_2$ are the natural projections and $\Poin$ is the
pullback of the Poincar\'e line bundle to
$M\times\torus^n\times\dtorus{}^n$ by projection. The (smooth)
Fourier-Mukai transform is then the isomorphism of K-theory
\be
T_!\,:\,\K^\bullet\big(M\times\torus^n\big)~\xrightarrow{\approx}~
\K^{\bullet+n}\big(M\times\dtorus{}^n\big)
\label{FMKtransf}\ee
given by
\be
T_!(-)=(p_2)_!\big(p_1^*(-)\otimes\Poin\big) \ .
\label{FMtransfexpl}\ee
It follows that \emph{topological T-duality} is a correspondence,
which may be described somewhat more explicitly as follows
(see~ref.~\cite{BMRS1} for more details).

By Rieffel's imprimitivity theorem, there is a Morita equivalence
\be
C_0(M\times\torus^n)\rtimes\R^n~\sim~C_0(M)\otimes C^*(\R^n)
\cong C_0\big(M\times\dtorus{}^n\big)
\label{Rieffelimthm}\ee
where the locally compact abelian group $\R^n$ acts trivially on $M$
and by translations on $\R^n/\Z^n$. By the Connes-Thom isomorphism,
this then defines a \emph{KK-equivalence}
\be
\alpha~\in~\KK_n\big(M\times\torus^n\,,\,M\times\dtorus{}^n\big) \ .
\label{alphaKKeq}\ee
The invertible element $\alpha$ may be interpreted analytically as the
families Dirac operator, and its inverse is the Bott element. It
follows that topological T-duality may be interpreted algebraically as
taking a crossed product with the natural action of $\R^n$ on the
$C^*$-algebra $C_0(M\times\torus^n)$. By Takai duality, there is a
Morita equivalence
\be
\big(C_0(M\times\torus^n)\rtimes\R^n\big)\rtimes\R^n\sim
C_0(M\times\torus^n)
\label{Takaidual}\ee
and hence the duality is of order two up to Morita equivalence.

These constructions all generalize~\cite{BMRS1,BMRS2} to the examples,
discussed in Section~\ref{Intro}, of D-branes in a spacetime $X$ which
is a principal torus bundle $X\to M$ in a constant $H$-flux
background. In this case D-branes are classified by the twisted
K-theory of $X$, defined as the K-theory of the continuous trace
algebra $\alg=CT(X,H)$. This is a noncommutative $C^*$-algebra with
spectrum equal to $X$ and Dixmier-Douady invariant equal to
$H\in\H^3(X,\Z)$. See ref.~\cite{BMRS2} for other noncommutative
examples of correspondences.

\subsection{Axiomatic T-duality}

The interpretations of T-duality in the examples considered in
Section~\ref{FourierMukai} above lead to the following general,
algebraic characterization of topological open string
T-duality. Consider a suitable category of separable $C^*$-algebras,
possibly equipped with some extra structure (such as the
$\R^n$-actions used in Section~\ref{FourierMukai} above), whose
objects $\alg$ are called \emph{T-dualizable algebras} and satisfy the
following properties:
\begin{enumerate}
\item There exists a covariant functor $\alg\mapsto T(\alg)$
  sending the algebra $\alg$ to the \emph{T-dual} of $\alg$;
\item There exists a functorial map
  $\alg\mapsto\alpha_\alg\in\KK(\alg,T(\alg))$ such that the class
  $\alpha_\alg$ is a KK-equivalence; and
\item The algebras $\alg$ and $T(T(\alg))$ are Morita
  equivalent, and the class
  $\alpha_\alg\otimes_{T(\alg)}\alpha_{T(\alg)}$ is the associated
  KK-equivalence.
\end{enumerate}
This realization of T-duality as a particular functorial kind of
involutive KK-equivalence gives a refinement of the more commonly used
notion of topological T-duality at the level of K-theory
alone~\cite{BMRS2}. This can be seen by noticing that, for
$C^*$-algebras $\alg,\balg$ which are KK-equivalent to commutative
$C^*$-algebras, there is a universal coefficient theorem presenting
the abelian group $\KK_\bullet(\alg,\balg)$ as an extension of
$\Hom_\Z(\K_\bullet(\alg),\K_\bullet(\balg))$ by
$\Ext_\Z(\K_{\bullet+1}(\alg),\K_\bullet(\balg))$~\cite{RS87}. The
characterization described here has the advantage that, unlike the
non-geometric examples such as T-folds, our T-dual spacetimes are
\emph{globally} defined, at the cost of possibly being
noncommutative.

\setcounter{equation}{0}
\section{D-brane charges on noncommutative spaces}

We will finally give the noncommutative versions of the
constructions of D-brane charges presented in
Section~\ref{DbranesK}. For this, we will first have to deal with
some topics of independent mathematical interest, which are concerned
with the problem of constructing certain topological invariants for
noncommutative spaces. In particular, we will describe a
noncommutative version of the Riemann-Roch theorem.

\subsection{Local cyclic cohomology}

To proceed further, we need to find an appropriate cohomological
analog of Kasparov's KK-functor. In particular, we need a bivariant
cohomology theory which is defined on a similar class of algebras as
KK-theory, which possesses similar algebraic and topological
properties, and which provides an appropriate receptacle for a suitably
defined Chern character. The best suited theory for our purposes is
Puschnigg's theory~\cite{Puschnigg} which can be defined on large
classes of topological and bornological algebras, as well as for
separable $C^*$-algebras. We will now briefly summarize the main
ingredients of this cyclic cohomology theory.

Let $\alg$ be a unital algebra, and let
$T\alg=\bigoplus_{n\geq0}\,\alg^{\otimes n}$ be the quasi-free tensor
algebra of $\alg$. On $T\alg$, define the algebra of
\emph{noncommutative differential forms} in degree $n$ by
\be
\Omega^n(\alg)=\alg^{\otimes(n+1)}\oplus\alg^{\otimes n} \ .
\label{Omeganalgdef}\ee
On the graded vector space $\Omega^\bullet(\alg)$ there is a
differential $\dd$ of degree~$+1$ defined on the splitting
(\ref{Omeganalgdef}) by
\be
\dd=\begin{pmatrix}0&0 \\ 1&0 \end{pmatrix}
\label{diffdef}\ee
and one has an isomorphism
\be
\Omega^n(\alg)\cong{\rm Span}_\C\big\{a_0~\dd a_1\cdots\dd a_n~
\big|~ a_0,a_1,\dots,a_n\in\alg\big\} \ .
\label{Omegandd}\ee
We are interested in a suitable completion of the algebra
$\Omega^\bullet(\alg)$ which can be described as a certain deformation
$X(T\alg)$ of the tensor algebra. This is the $\Z_2$-graded
\emph{$X$-complex} defined by
\be
\xymatrix{
X(T\alg)\,:\,\Omega^0(T\alg)=T\alg~
\ar@<0.5ex>[rr]^{\!\!\!\!\!\!\!\!\!\!\!\!\!\!\!\!
\!\!\!\!\!\!\!\!\!\!\!\natural\circ\dd} &&
\ar@<0.5ex>[ll]^{\!\!\!\!\!\!\!\!\!\!\!\!\!\!\!\!
\!\!\!\!\!\!\!\!\!\!\!\bb}~\Omega^1(T\alg)_\natural=
\Omega^1(T\alg)\,{\big/}\,
\big[\Omega^1(T\alg)\,,\,\Omega^1(T\alg)\big] \ ,
}
\label{Xcomplex}\ee
where $\bb$ is the nilpotent operator defined by
$\omega_0~\dd\omega_1\mapsto[\omega_0,\omega_1]$ for
$\omega_0,\omega_1\in T\alg$ and $\natural$ denotes the quotient map
$\Omega^1(T\alg)\to\Omega^1(T\alg)_\natural$.

With some additional structure~\cite{BMRS1}, one can then define the
$\Z_2$-graded \emph{bivariant local cyclic cohomology}
\be
\HL_\bullet(\alg,\balg)=\H_\bullet\big(\Hom_\C(\,
\widehat{X}(T\alg),\widehat{X}(T\balg)\,)\,,\,\partial\big)
\label{HLdef}\ee
where $\partial$ is a differential determined by the $X$-complex
(\ref{Xcomplex}) which makes
$\Hom_\C(\widehat{X}(T\alg),\widehat{X}(T\balg))$ into a
$\Z_2$-graded complex of bounded maps, and
\be
\xymatrix{\displaystyle
\widehat{X}(T\alg)\,:\,\prod_{n\geq0}\,\Omega^{2n}(\alg)~
\ar@<1.0ex>[rr] && \ar@<1.0ex>[ll]~\displaystyle\prod_{n\geq0}\,
\Omega^{2n+1}(\alg)
}
\label{Puschniggcompl}\ee
is the Puschnigg completion of $X(T\alg)$. This bivariant cyclic
cohomology theory is the closest one in structure to Kasparov's
KK-theory. In particular, most of our previous definitions and
constructions in bivariant K-theory have obvious analogs in local
bivariant cyclic cohomology. The key property of this theory is the
existence of a ``good'' $\Z_2$-graded bivariant Chern character map
\be
\ch\,:\,\KK(\alg,\balg)~\longrightarrow~\HL(\alg,\balg)
\label{chbivariant}\ee
for any two separable $C^*$-algebras $\alg$ and $\balg$. The
homomorphism (\ref{chbivariant}) is functorial and multiplicative.

As an explicit example, let $X$ be a compact oriented manifold of
dimension~$d$. Then the inclusion
\be
C^\infty(X)~\hookrightarrow~C(X)
\label{Frechetincl}\ee
of the Fr\'echet algebra of smooth functions on $X$ determines an
invertible element of $\HL(C^\infty(X),C(X))$~\cite{BMRS1}, and hence
an HL-equivalence. Thus $\HL(C(X))\cong\HL(C^\infty(X))$, for both
homology and cohomology. On the other
hand, $\HL(C^\infty(X))$ is isomorphic to the standard periodic cyclic
homology $\HP(C^\infty(X))$~\cite{Puschnigg}. Moreover, the action of
the boundary map $\bb$ in eq.~(\ref{Xcomplex}) is trivial in this case
and the Puschnigg complex (\ref{Puschniggcompl}) reduces to the
periodic complexified de~Rham complex $(\Omega^\bullet(X),\dd)$, where
$\dd$ is the usual de~Rham exterior derivative on $X$. Connes' version
of the Hochschild-Kostant-Rosenberg theorem gives a quasi-isomorphism
$\mu:\Omega^n(C^\infty(X))\to\Omega^n(X)$ which is implemented by
sending a noncommutative $n$-form to a differential $n$-form,
\be
\mu\big(f^0~\dd f^1\cdots\dd f^n\big)=\mbox{$\frac1{n!}$}~f^0~\dd
f^1\wedge\cdots\wedge\dd f^n \ , \quad f^i\in C^\infty(X) \ .
\label{canquasimu}\ee
It follows that the periodic cyclic homology of the
algebra $C^\infty(X)$ is canonically isomorphic to the periodic
de~Rham cohomology of $X$. Putting everything together we thus have
the $\Z_2$-graded isomorphism
\be
\HL_\bullet\big(C(X)\big)\cong\H_{\rm dR}^\bullet(X) \ .
\label{HLHdRiso}\ee
Moreover, the image of the class $[\varphi]$ of the cyclic
$d$-cocycle
\be
\varphi\big(f^0\,,\,f^1\,,\, \dots\,,\, f^d\big) = \frac1{d!}\,
\int_X\, f^0~\dd f^1\wedge\dots\wedge \dd f^d
\label{varphiX}\ee
under the homomorphism $\HP^\bullet(C^\infty(X))\cong
\HL^\bullet(C(X))\to\HL^\bullet(C(X)\otimes C(X))$ induced by the
product map is the orientation fundamental class
$\Xi\in\HL^d(C(X)\otimes C(X))$ of $X$ in cyclic cohomology. Higher
degree homology classes of $X$ are obtained by associating in this way
a cyclic $k$-cocycle with any closed $k$-current $C$ on $X$.

\subsection{Todd classes}

Let $\alg$ be a PD algebra with fundamental K-homology class
$\Delta\in\K^d(\alg\otimes\alg^\op)$ and fundamental cyclic cohomology
class $\Xi\in\HL^d(\alg\otimes\alg^\op)$. Then the \emph{Todd class}
of $\alg$ is the element of the unital ring $\HL_0(\alg,\alg)$ given
by
\be
\Todd(\alg):=\Xi^\vee\otimes_{\alg^\op}\ch(\Delta) \ .
\label{Todddef}\ee
The Todd class depends ``covariantly'' on the choices of fundamental
classes~\cite{BMRS1}, and it is invertible with inverse class given by
\be
\Todd(\alg)^{-1}=(-1)^d~\ch\big(\Delta^\vee\big)\otimes_{\alg^\op}
\Xi \ .
\label{Toddinv}\ee
This definition can be motivated by the following example. Let
$\alg=C(X)$ where $X$ is a compact complex manifold. Then $\alg$ is a
PD algebra with KK-theory fundamental class $\Delta$ given by the
Dolbeault operator $\partial$ on $X\times X$, and with HL-theory
fundamental class $\Xi$ induced by the orientation cycle $[X]$
determining Poincar\'e duality in rational homology of $X$. As
before, there is an isomorphism $\HL(\alg)\cong\HP(C^\infty(X))$, and
so by the universal coefficient theorem we may identify
$\HL(\alg,\alg)\cong\End(\H(X,\Q))$. Then
$\Todd(\alg)=\,\smile\,\Todd(X)$ is cup product with the usual Todd
characteristic class $\Todd(X)\in\H(X,\Q)$.

\subsection{Grothendieck-Riemann-Roch theorem}

For any K-oriented morphism $f:\alg\to\balg$ of separable
$C^*$-algebras, we can compare the bivariant cyclic cohomology
class $\ch(f!)$ with the HL-theory orientation class $f*$ in
$\HL_d(\balg,\alg)$. If $\alg$ and $\balg$ are both PD algebras, then
$d=d_\alg-d_\balg$ and one has the noncommutative
\emph{Grothendieck-Riemann-Roch formula}
\be
\ch(f!)=(-1)^{d_\balg}~\Todd(\balg)\otimes_\balg(f*)\otimes_\alg
\Todd(\alg)^{-1} \ .
\label{GRRformula}\ee
This may be proven by writing out both sides of eq.~(\ref{GRRformula})
using the various definitions and multiplicativity properties of the
bivariant Chern character, and then simplifying using associativity
properties of the intersection product~\cite{BMRS1,BMRS2}. In terms of
the associated Gysin maps, the formula (\ref{GRRformula}) yields a
commutative diagram
\be
\xymatrix{
\K_\bullet(\balg)~\ar[r]^{\!\!f_!}
\ar[d]_{\ch\otimes_\balg\Todd(\balg)} & 
~\K_{\bullet+d}(\alg)\ar[d]^{\ch\otimes_\alg\Todd(\alg)} \\
\HL_\bullet(\balg)~\ar[r]_{\!\!\!\!\!f_*} & ~
\HL_{\bullet+d}(\alg) \ . 
}
\label{GRRcommdiag}\ee
See ref.~\cite{BMRS2} for some applications of this theorem.

\subsection{Isometric pairing formula}

Let $\alg$ be a PD algebra. A fundamental K-homology class $\Delta$
for $\alg$ is called \emph{symmetric} if $\sigma(\Delta)^\op=\Delta$ in
$\K^d(\alg\otimes\alg^\op)$, where $\sigma$ is the map on K-homology
induced by the flip involution
$\alg\otimes\alg^\op\to\alg^\op\otimes\alg$ defined by $x\otimes
y^\op\mapsto y^\op\otimes x$ for $x,y\in\alg$. If $\alg$ satisfies the
universal coefficient theorem for local bivariant cyclic cohomology,
then $\HL_\bullet(\alg,\alg)\cong\End(\HL_\bullet(\alg))$. If
$\HL_\bullet(\alg)$ is a finite-dimensional vector space, then the
bivariant Chern character (\ref{chbivariant}) induces an isomorphism
\be
\ch\,:\,\KK(\alg,\balg)\otimes_\Z\C\cong
\Hom_\C\big(\K_\bullet(\alg)\otimes_\Z\C \,,\,
\K_\bullet(\balg)\otimes_\Z\C\big)~\xrightarrow{\approx}~
\HL(\alg,\balg)
\label{chbiiso}\ee
for any separable $C^*$-algebra $\balg$ which is KK-equivalent to a
commutative $C^*$-algebra. If in addition $\alg$ has symmetric
(even-dimensional) cyclic and K-theory fundamental classes $\Xi$ and
$\Delta$, then the \emph{modified Chern character}
\be
\ch\otimes_\alg\sqrt{\Todd(\alg)}\,:\,\K_\bullet(\alg)~
\longrightarrow~\HL_\bullet(\alg)
\label{modchNC}\ee
is an isometry with respect to the inner products
\be
(\alpha,\beta)_\K=\big(\alpha\otimes\beta^\op\big)
\otimes_{\alg\otimes\alg^\op}\Delta
\label{NCKinnprod}\ee
on the K-theory of $\alg$ and
\be
(x,y)_\HL=\big(x\otimes y^\op\big)\otimes_{\alg\otimes\alg^\op}\Xi
\label{NCHLinnprod}\ee
on the local cyclic homology of $\alg$.

The proof of this result can be found in refs.~\cite{BMRS1,BMRS2}. Let
us comment on some of the ingredients that go into this formula. The
assumptions made on the local cyclic homology of $\alg$ enable one to
identify the Todd class $\Todd(\alg)$ as an element of
$GL(\HL_0(\alg))\cong GL(n,\C)$, where $n:=\dim_\C(\HL_0(\alg))$. One
can then define its square root using the Jordan canonical form, and
then consider $\sqrt{\Todd(\alg)}$ as an element of the ring
$\HL_0(\alg,\alg)$ again by using the universal coefficient
theorem. This square root is not unique, but we fix a choice. In some
instances the Todd class may be self-adjoint and positive with
respect to an inner product on the vector space $\HL_0(\alg)$, which
may help to fix a canonical choice. The symmetry hypothesis made on
the fundamental classes of $\alg$ ensure that the pairings
(\ref{NCKinnprod}) and (\ref{NCHLinnprod}) define symmetric bilinear
forms. Then, by using the definition of the intersection product, one
can show that the pairing (\ref{NCKinnprod}) generalizes the index
pairing (\ref{Kpairing}) on topological K-theory. Indeed, in the
commutative case, this result is essentially the KK-theory version of
the Atiyah-Singer index theorem.

Using the modified Chern character (\ref{modchNC}), we finally arrive
at a noncommutative version of the Minasian-Moore formula
(\ref{MMformula}) valued in $\HL_\bullet(\alg)$ and given by
\be
Q(\dalg,\xi)=\ch\big(f_!(\xi)\big)\otimes_\alg\sqrt{\Todd(\alg)} \ ,
\label{NCMMformula}\ee
for noncommutative D-branes $\alg,\dalg$ with K-oriented morphism
$f:\alg\to\dalg$ and Chan-Paton bundle $\xi\in\K_\bullet(\dalg)$. By
using the noncommutative Grothendieck-Riemann-Roch formula
(\ref{GRRformula}), one can formulate criteria under which the charge
vector (\ref{NCMMformula}) can be expressed as the pullback $f_*$ of a
``Wess-Zumino class'' in $\HL_\bullet(\dalg)$ and hence formulate
conditions under which the noncommutative D-brane charge is invariant
under T-duality transformations. See ref.~\cite{BMRS2} for an explicit
description of this and for applications of the noncommutative charge
formula (\ref{NCMMformula}).

\section*{Acknowledgments}

R.J.S. would like to thank the organizers of the Alessandria and Orsay
meetings for the invitation and hospitality in a stimulating
atmosphere. He also thanks P.~Bieliavsky, P.~Turner and S.~Willerton
for helpful comments. V.M.\ was supported by the Australian Research
Council. J.R.\ was partially supported by US NSF grant
DMS-0504212. R.J.S.\ was supported in part by the EU-RTN Network Grant
MRTN-CT-2004-005104.

\end{document}